\documentclass[12pt,twoside]{article}

\evensidemargin=\oddsidemargin
\def\Title#1#2#3{%
    \baselineskip=18pt
    \begin{center}
          {\large\bf\uppercase{#1} \\ }
          \bigskip\bigskip
          {#2} \\
          {#3} \\
    \end{center}}
\long\def\Abstract#1{%
         \bigskip
         \parbox{0.93\textwidth}{%
                 \begin{center}
                       {\bf Abstract} \\
                 \end{center}
                 \medskip{\baselineskip=14pt #1}
                 \vss}
         \bigskip}

\makeatletter
\renewcommand{\section}%
 {\@startsection{section}{1}{0pt}%
  {-3.25ex plus -1ex minus -.2ex}{1.5ex plus .2ex}%
  {\vspace*{5mm}\raggedright\large\bf }}
\renewcommand{\thesection}{\arabic{section}.}
\@addtoreset{equation}{section}
\renewcommand{\@eqnnum}{(\thesection\theequation)}
\renewcommand{\p@equation}{\thesection}
\makeatother

\begin{document}

\vspace*{1cm}

\Title{THE ``EXTENDED PHASE SPACE'' APPROACH\\
TO QUANTUM GEOMETRODYNAMICS:\\
WHAT CAN IT GIVE FOR THE DEVELOPMENT\\
OF QUANTUM GRAVITY?}%
{T. P. Shestakova}%
{Department of Theoretical and Computational Physics,
Southern Federal University\footnote{former Rostov State University},\\
Sorge St. 5, Rostov-on-Don 344090, Russia \\
E-mail: {\tt shestakova@phys.rsu.ru}}

\Abstract{The talk is devoted to the ``extended phase space'' approach to Quantum Geometrodynamics. The premises that have led to the formulation of this approach are briefly reviewed, namely, non-trivial topology of the Universe which implies the absence of asymptotic states, in contrast to situations one usually deals in ordinary quantum field theory; parametrization noninvariance in the Wheeler -- DeWitt theory; the problem of time and the absence of dynamical evolution. Then we discuss the main features of the approach: Hamiltonian dynamics in extended phase space, gauge-dependent Schr\"odinger equation for the wave function of the Universe, the description of quantum Universe from the viewpoint of observers in a wide enough class of reference frames. After all, we analyse problems arising in this approach: the structure of Hilbert space in Quantum Geometrodynamics, the relations between solutions for the wave function of the Universe corresponding to various reference frames, properties of a medium to be necessary to fix a reference frame, the transition to classical limit.}

\section{Introduction}
The purpose of my talk is to present a new approach, the so-called ``extended phase space'' approach to Quantum Geometrodynamics, which was proposed by the group of researchers, G. M. Vereshkov, V. A. Savchenko and me, from Rostov State University (the University has been rearranged into Southern Federal University, and sometimes it arouses misunderstanding) in the end of 1990s \cite{SSV1,SSV2,SSV3,SSV4}. A search for a new approach was inspired by the well-known problems of the Wheeler -- DeWitt quantum geometrodynamics such as the problem of time and the problem of reparametrization noninvariance. The most of the paper on quantum geometrodynamics comprised attempts to solve these problems in the limits of the Wheeler -- DeWitt quantum geometrodynamics without a thorough analysis of their origin. On the other hand, quite radial opinions have also appeared, as the one by Isham, who wrote \cite{Isham}:  ``...although it may be heretical to suggest it, the Wheeler -­ DeWitt equation -- elegant though it be -- may be completely the wrong way of formulating a quantum theory of gravity''.

To my mind, the Wheeler -- DeWitt quantum geometrodynamics, the first significant attempt to construct full quantum theory of gravity \cite{DeWitt}, is based on three cornerstones: Dirac approach to quantization of systems with constraints \cite{Dirac1,Dirac2}, Arnowitt -– Deser -– Misner (ADM) parametrization \cite{ADM} and the ideas of Wheeler concerning a wave functional describing a state of gravitational field \cite{Wheeler1,Wheeler2}. Let me characterize very briefly each of these point.

In the Dirac approach the central part is given to a postulate, according to which each constraint $\varphi_m(q,p)=0$ after quantization becomes a condition on a state vector, or wave functional, $\Psi$:
\begin{equation}
\label{Dir.cond}
\varphi_m\Psi=0.
\end{equation}
Let us emphasize that it is indeed a postulate, since it cannot be justified by the reference to the correspondence principle. The role prescribed to the constraints could be explained by the fact that at the classical level, the constraints express gauge invariance of the theory. It was initially believed that imposing constraints at the quantum level would also ensure gauge invariance of wave functional. {\it But what grounds do we have to expect it?} Strictly speaking, this issue have not been investigated and gauge invariance of the theory has not been proved. It leads us to the next fundamental problem: Could we consider quantum geometrodynamics as a gauge-invariant theory?

An important role was played by the ADM parametrization which has a clear geometrical interpretation, and it is the ADM parametrization that enables one to write gravitational constraints in the form independent of gauge variables -- the lapse and shift functions $N$, $N_i$. It gave rise to an illusion that the theory in which the main equations are those of constraints must not depend on a choice of gauge conditions. At the same time, as was emphasized in the works of the group of Montani, the ADM parametrization introduces in 4-dimensional spacetime (3+1)-splitting, fixing (3+1)-splitting prescribes particular values for the lapse and shift functions \cite{MM1,MM2} that is equivalent to a choice of a reference frame, and gauge invariance breaks down. Thus, {\it the Hamiltonian constraint loses its sense and, with the latter, so does the whole procedure of quantization}.

The third point was the idea by Wheeler that the wave functional must be determined on the superspace of all possible 3-geometries. However, the statement that the wave function must depend only on 3-geometry is just a declaration without any mathematical realization. As we know, the state vector always depends on a concrete form of the metric.

Here is my assessment of the Wheeler -- DeWitt theory as I see it at present. In the beginning of our work we were seeking for an approach which would enable one to analyse mathematical correctness and consistency of constructing quantum geometrodynamics. We considered the Wheeler -- DeWitt theory \cite{DeWitt} as the extrapolation of conceptions and methods of modern quantum field theory, the validity of which on the scale of the Universe as a whole may arouse doubts.

\section{The premises for the ``extended phase space'' approach to Quantum Geometrodynamics}
Gauge invariance of the Wheeler -- DeWitt theory can hardly be investigated within canonical quantization approach. Path integration approach is more powerful: it contains the procedure of derivation of an equation for a wave function from the path integral, while gauge invariance of the path integral, and the theory as a whole, being ensured by asymptotic boundary conditions. In ordinary quantum theory one usually considers systems with asymptotic states in which the so-called physical and non-physical degrees of freedom could be separated from each other. Asymptotic boundary conditions in the path integral are equivalent to selection rules for physical states. The only case of a gravitating system with asymptotic states is the case of asymptotically flat spacetime. One of the main premises of our approach was the fact that a universe with non-trivial topology, in particular, a closed universe, does not possess asymptotic states. It is worth mentioning that in first works devoted to derivation of the Wheeler -- DeWitt equation from the path integral \cite{BP,Hall}, asymptotic boundary conditions were tacitly adopted without careful consideration if they are justified. We formulated the purpose of our work in another way: since we were not sure that in the absence of asymptotic states in a topologically non-trivial universe we would be able to construct a gauge-invariant theory, we had no grounds at all to require for a wave function to satisfy the Wheeler -- DeWitt equation. At the same time, independently on our notion about gauge invariance or noninvariance of the theory, the wave function has to obey some Schr\"odinger equation. Only after constructing the wave function satisfying the Schr\"odinger equation, we shall be able to investigate the question, if this wave function obey the Wheeler -- DeWitt equation as well.

The second premise of our approach was parametrization noninvariance in the Wheeler -- DeWitt theory. The parametrization noninvariance is the well-known fact, however, we demonstrated that it implies a hidden gauge noninvariance. Indeed, the choice of gauge variables and the choice of gauge conditions have a unified interpretation: they together determine equations for the metric components $g_{0\mu}$, fixing a reference frame.

\begin{center}
\begin{tabular}{ccccc}
Parametrization & + & Gauge conditions & $\Rightarrow$ &
 Equations for $g_{0\mu}$\\
$g_{0\mu}=v_{\mu}\left(\tilde N_{\nu},\gamma_{ij}\right)$ & &
 $\tilde N_{\nu}=f_{\nu}\left(\gamma_{ij}\right)$ & &
 $g_{0\mu}=v_{\mu}\left(f_{\nu}\left(\gamma_{ij}\right),\gamma_{ij}\right)$
\end{tabular}
\end{center}

\noindent
Here $\tilde N_{\nu}$ are new gauge variables, in particular, the lapse and shift functions, $N$ and $N_i$, $\gamma_{ij}$ is 3-metric. Thus even if one considers $\tilde N_{\nu}$ as independent of $\gamma_{ij}$, different parametrizations will correspond to different reference frames. One may think that we do not need any conditions on gauge variables $\tilde N_{\nu}$ if the constraints seem not to depend on them. Actually, to define the operator form of the constraints after quantization (or, in other words, to solve the ordering problem), we do need to know the relations between the gauge variables and the rest ones. The first who pointed to this fact were Hawking and Page \cite{HP}. It is important to understand that the ordering problem cannot be solved without making use, explicitly or implicitly, of the additional condition on $N$. This additional relations can play the role of gauge condition.

I shall not talk in detail about the fundamental problems of the Wheeler -- DeWitt theory which all are well-known. They are: the problem of time (the absence of dynamical evolution), the problem of Hilbert space, the problem of observables and others. Let me emphasize that all these problems are interrelated, the problem of time creates that of Hilbert space, etc.

\section{The choice of quantization scheme}
Let me turn to the description of our approach to construction of quantum geometrodynamics. As I have already said above, the path integral approach is more powerful for analyzing the procedure of derivation of an equation for a wave function. According to the physical situation, we consider the path integral without asymptotic boundary conditions.

The path integral formalism does not require to construct Hamiltonian form of the theory at all since the equation for a wave function can be derived directly from a path integral in Lagrangian form. However, we can still choose between two formulations: we can deal with the path integral in Lagrangian form with Batalin -- Vilkovisky (in fact, Faddeev -- Popov) effective action, or deal with the path integral in Hamiltonian form with Batalin -- Fradkin -- Vilkovisky effective action with following integrating out all momenta and passing on to a path integral over extended configurational space. I would like to point to the importance of this choice which actually is the choice between two different theories. Indeed, there exists the difference between the group of transformations generated by gravitational constrains in Hamiltonian formalism and that of gauge transformations of the Einstein theory (in Lagrangian formalism). Already at the classical level we deal with two different theories of gravity which are invariant under different groups of transformations in Lagrangian and Hamiltonian formulations. In the path integral quantization these transformations define the structure of ghost sectors which also appear to be different. The two formulations could enter into agreement only in a gauge-invariant sector which can be singled out by asymptotic boundary conditions; the later ones must supposedly pick out trivial solutions for ghosts and Lagrange multipliers to ensure gauge-invariant dynamics.

However, if one consider the Universe as a system which, in general, does not possess asymptotic states, one have to pose the questions:

1. What formalism should one prefer?

2. What are consequences of the fact that we consider the path integral without asymptotic boundary conditions?

3. What would be a role of gauge degrees of freedom, which were traditionally considered as redundant, in this new approach?

\section{The model}
The answer for the first question is: Our choice was the Batalin -- Vilkovisky (Lagrangian) formalism which corresponds to the original (Einstein) formulation of gravitational theory. It is convenient to illustrate our approach for a simple cosmological model with a finite number of degrees of freedom. The action for the model reads
\begin{equation}
\label{action}
S=\!\int\!dt\,\biggl\{
  \displaystyle\frac12 v(\tilde N, Q)\gamma_{ab}\dot{Q}^a\dot{Q}^b
  -\frac1{v(\tilde N, Q)}U(Q)
  +\pi_0\left(\dot{\tilde N}-f_{,a}\dot{Q}^a\right)
  -i w(\tilde N, Q)\dot{\bar\theta}\dot\theta\biggr\}.
\end{equation}
Here $Q=\{Q^a\}$ stands for physical variables such as a scale factor or gravitational-wave degrees of freedom and material fields, and we use an arbitrary parametrization of a gauge variable $\tilde N$ determined by the function $v(\tilde N, Q)$. In the case of isotropic universe or the Bianchi IX model $\tilde N$ is bound to the scale factor $a$ and the lapse function $N$ by the relation
\begin{equation}
\label{paramet}
\displaystyle\frac{a^3}{N}=v(\tilde N, Q).
\end{equation}
\begin{equation}
\label{w_def}
w(\tilde N, Q)=\frac{v(\tilde N, Q)}{v_{,\tilde N}};\quad
v_{,\tilde N}\stackrel{def}{=}\frac{\partial v}{\partial\tilde N}.
\end{equation}
$\theta,\,\bar\theta$ are the Faddeev -- Popov ghosts after replacement $\bar\theta\to -i\bar\theta$. We work in the class of gauges not depending on time
\begin{equation}
\label{frame_A}
\tilde N=f(Q)+k;\quad
k={\rm const},
\end{equation}
which can be presented in a differential form,
\begin{equation}
\label{diff_form}
\dot{\tilde N}=f_{,a}\dot{Q}^a,\quad
f_{,a}\stackrel{def}{=}\frac{\partial f}{\partial Q^a}.
\end{equation}

Though we do not need the Hamiltonian formulation of the theory to derive a Schr\"odinger equation from the path integral, the differential form of gauge conditions enables us to construct the Hamiltonian in a usual way, according to the rule $H=P\dot Q-L$, where $(P,Q)$ are the canonical pairs of extended phase space (EPS), by introducing momenta conjugate to all degrees of freedom including the gauge and ghost ones,
\begin{equation}
\label{Hamilt}
H=P_a\dot Q^a+\pi_0\dot{\tilde N}+\bar\rho\dot\theta+\dot{\bar\theta}\rho-L
 =\frac12G^{\alpha\beta}P_{\alpha}P_{\beta} +\frac1{v(\tilde N,Q)}U(Q)
 -\frac i{w(\tilde N,Q)}\bar\rho\rho,\nonumber\\
\end{equation}
where $\alpha=(0,a),\;Q^0=\tilde N$,
\begin{equation}
\label{Galphabeta}
G^{\alpha\beta}=\frac1{v(\tilde N,Q)}\left(
\begin{array}{cc}
f_{,a}f^{,a}&f^{,a}\\
f^{,a}&\gamma^{ab}
\end{array}
\right).
\end{equation}
The Lagrange multiplier $\pi_0$ plays the role of the momentum conjugata to the only gauge variable $\tilde N$.

Varying the effective action (\ref{action}) with respect to $Q^a$, $\tilde N$, $\pi_0$ and $\theta$, $\bar\theta$ one gets, correspondingly, motion equations for physical variables, the constraint, the gauge condition and equations for ghosts. The extended set of Lagrangian equations is complete in the sense that it enables one to formulate the Cauchy problem. The explicit substitution of trivial solutions for ghosts and the Lagrangian multiplier $\pi_0$ to this set of equations turns one back to the gauge-invariant classical Einstein equations.

It is not difficult to check that the system of Hamiltonian equations in EPS
\begin{equation}
\label{Hamilt.Eqs.}
\dot{P} = - \frac{\partial H}{\partial Q};\quad
\dot{Q} = \frac{\partial H}{\partial P}
\end{equation}
is completely equivalent to the extended set of Lagrangian equations, the constraint and the gauge condition acquiring the status of Hamiltonian equations. The idea of extended phase space is exploited in the sense that gauge and ghost degrees of freedom are treated on an equal basis with other variables. This gave rise to the name ``quantum geometrodynamics in extended phase space''.

Since the Hamiltonian dynamics in EPS is completely equivalent to Lagrangian dynamics, the group of transformations in EPS corresponds to the group of gauge transformations in the Lagrangian formalism. One can construct the BRST generator,
\begin{equation}
\label{BRSTgen}
\Omega
 =w(Q,\tilde N)\;\pi_0\dot\theta-H\theta
 =-\;i\;\pi_0\rho-H\theta.
\end{equation}
It is easy to check that (\ref{BRSTgen}) generates transformations in EPS which are identical to the BRST transformations in the Lagrangian formalism. On the other hand, we have a prescription given by Batalin, Fradkin and Vilkovisky how to construct the BRST generator in Hamiltonian Formalism when we have the set of constraints ${\cal G}_{\alpha}=(\pi_0,\;{\cal T})$, $\pi_0$ being a primary constraint, and ${\cal T}$ is the Hamiltonian (secondary) constraint of the theory. The generator (\ref{BRSTgen}) does not coincide with the one constructed according to prescriptions by BFV:
\begin{equation}
\label{Omega_BFV}
\Omega_{BFV}=\eta^{\alpha}{\cal G}_{\alpha}={\cal T}\theta-i\pi_0\rho,
\end{equation}
In the BFV approach the Wheeler -- DeWitt equation ${\cal T}\,|\Psi\rangle=0$ immediately follows from the requirement of BRST invariance $\Omega_{BFV}\,|\Psi\rangle=0$ due to arbitrariness of BFV ghosts $\{\eta^{\alpha}\}$.

Because of the difference in groups of transformations the BFV charge (\ref{Omega_BFV}) turns our to be irrelevant in this consideration. At the same time, the ``new'' BRST generator (\ref{BRSTgen}) cannot be presented as a combination of constraints and does not lead to the Wheeler -- DeWitt equation.

This makes us to look in a new light at the status of BRST invariance. We have a theory in EPS which, after imposing a gauge condition, is still invariant under global BRST transformation. However, this simple example shows that after quantization of the theory the requirement of BRST invariance is not, in general, a remedy to restore the broken gauge invariance.

\section{The general solution to the Schr\"odinger equation and the role of gauge degrees of freedom}
Let us turn now to the quantization procedure. We derive the Schr\"odinger equation from the path integral with the effective action (\ref{action}) and without asymptotic boundary conditions by a standard method originated by Feynman \cite{Fey,Cheng}. For the present model it reads
\begin{equation}
\label{SE1}
i\,\frac{\partial\Psi(\tilde N,Q,\theta,\bar\theta;\,t)}{\partial t}
 =H\Psi(\tilde N,\,Q,\,\theta,\,\bar\theta;\,t),
\end{equation}
where
\begin{equation}
\label{H}
H=-\frac i w\frac{\partial}{\partial\theta}
   \frac{\partial}{\partial\bar\theta}
  -\frac1{2M}\frac{\partial}{\partial Q^{\alpha}}MG^{\alpha\beta}
   \frac{\partial}{\partial Q^{\beta}}+\frac1v(U-V[f]);
\end{equation}
the operator $H$ corresponds to the Hamiltonian in EPS (\ref{Hamilt}). This is another argument in favor of our choice of quantization scheme.

$M$ is the measure in the path integral,
\begin{equation}
\label{M}
M(\tilde N, Q)=v^{\frac K2}(\tilde N, Q)w^{-1}(\tilde N, Q);
\end{equation}
$K$ is a number of physical degrees of freedom; the wave function is defined on extended configurational space with the coordinates
$\tilde N,\,Q^a,\,\theta,\,\bar\theta$.
$V$ is a quantum correction to the potential $U$, that depends on the chosen parametrization (\ref{paramet}) and gauge (\ref{frame_A}). Its explicit form is the following:
\begin{eqnarray}
V[f]&=&\frac5{12w^2}\left(w^2_{,\mu}f_{,a}f^{,a}+2w_{,\mu}f_{,a}w^{,a}
    +w_{,a}w^{,a}\right)
   +\frac1{3w}\left(w_{,\mu,\mu}f_{,a}f^{,a}+2w_{,\mu,a}f^{,a}
    +w_{,\mu}f_{,a}^{,a}+w_{,a}^{,a}\right)+\nonumber\\
&+&\frac{K-2}{6vw}\left(v_{,\mu}w_{,\mu}f_{,a}f^{,a}+v_{,\mu}f_{,a}w^{,a}
    +w_{,\mu}f_{,a}v^{,a}+v_{,a}w^{,a}\right)-\nonumber\\
&-&\frac{K^2-7K+6}{24v^2}\left(v^2_{,\mu}f_{,a}f^{,a}+2v_{,\mu}f_{,a}v^{,a}
    +v_{,a}v^{,a}\right)+\nonumber\\
\label{V}
&+&\frac{1-K}{6v}\left(v_{,\mu,\mu}f_{,a}f^{,a}+2v_{,\mu,a}f^{,a}
    +v_{,\mu}f_{,a}^{,a}+v_{,a}^{,a}\right).
\end{eqnarray}

The general solution to the Schr\"odinger equation has the following structure:
\begin{equation}
\label{GS-A}
\Psi(\tilde N,\,Q,\,\theta,\,\bar\theta;\,t)
 =\int\Psi_k(Q,\,t)\,\delta(\tilde N-f(Q)-k)\,(\bar\theta+i\theta)\,dk.
\end{equation}
It is a superposition of eigenstates of a gauge operator,
\begin{equation}
\label{k-vector}
\left(\tilde N-f(Q)\right)|k\rangle=k\,|k\rangle;\quad
|k\rangle=\delta\left(\tilde N-f(Q)-k\right).
\end{equation}
It can be interpreted in the spirit of Everett's ``relative state'' formulation. In fact, each element of the superposition (\ref{GS-A}) describe a state in which the only gauge degree of freedom $\tilde N$ is definite, so that time scale is determined by processes in the physical subsystem through functions $v(\tilde N,\,Q),\,f(Q)$ (see (\ref{paramet}), (\ref{frame_A})), while $k$ being determined by initial clock setting. Indeed, according to (\ref{frame_A}), the parameter $k$ gives an initial condition for the variable $\tilde N$. The function $\Psi_k(Q,\,t)$ describes a state of the physical subsystem for a reference frame fixed by the condition (\ref{frame_A}). It is a solution to the equation
\begin{equation}
\label{phys.SE}
i\,\frac{\partial\Psi_k(Q;\,t)}{\partial t}
 =H_{(phys)}[f]\Psi_k(Q;\,t),
\end{equation}
\begin{equation}
\label{phys.H-A}
H_{(phys)}[f]=\left.\left[-\frac1{2M}\frac{\partial}{\partial Q^a}
  \frac1v M\gamma^{ab}\frac{\partial}{\partial Q^b}
 +\frac1v (U-V)\right]\right|_{\tilde N=f(Q)+k}.
\end{equation}

The dependence of $\Psi_k(Q,\,t)$ on $k$ is not fixed by the
equation (\ref{phys.SE}) in the sense that $\Psi_k(Q,\,t)$ can be
multiplied by an arbitrary function of $k$. On the other side, one
cannot choose the function $\Psi_k(Q,\,t)$ to be not depending on
$k$, since in this case one would obtain a non-normalizable,
non-physical state. The normalization condition for the wave
function (\ref{GS-A}) reads
$$
\int\Psi^*(\tilde N,\,Q,\,\theta,\,\bar\theta;\,t)\,
 \Psi(\tilde N,\,Q,\,\theta,\,\bar\theta;\,t)\,M(\tilde N,\,Q)\,
 d\tilde N\,d\theta\,d\bar\theta\,\prod_adQ^a=
$$
$$
\int\Psi^*_k(Q,\,t)\,\Psi_{k'}(Q,\,t)\,
 \delta(\tilde N-f(Q)-k)\,\delta(\tilde N-f(Q)-k')\,M(\tilde N,\,Q)\,
 dk\,dk'\,d\tilde N\,\prod_adQ^a=
$$
\begin{equation}
\label{Psi_norm}
=\int\Psi^*_k(Q,\,t)\,\Psi_k(Q,\,t)\,
 M(f(Q)+k,\,Q)\,dk\,\prod_adQ^a=1.
\end{equation}

By introducing a certain gauge condition we determine a gauge subsystem of the Universe which affects properties of physical Universe. The gauge subsystem shows itself as a real constituent of the Universe. Indeed, firstly, a chosen gauge condition determines the form of the equation
(\ref{phys.SE}) for the physical part of the wave function $\Psi_k(Q,\,t)$, in particular, an effective quantum potential. Secondly, the measure $M\left(f(Q)+\bar k,\,Q\right)$ in physical subspace also depends on the gauge condition, so that any changes of the gauge condition result in changes of the measure. In other words, if we determined the gauge subsystem in some different way, it would reflect on the structure of physical subspace.

Here is the answer for the second question: the gauge dependent Schr\"odinger equation and the structure of the wave function are direct consequences of the fact that we consider the path integral without asymptotic boundary conditions.

But what is the gauge subsystem? Let us return to the extended set of equations obtained by varying this effective action. This set of equations includes ghosts equations and a gauge condition, and equations for physical degrees of freedom also contain gauge-noninvariant terms. So, the gauged Einstein equations look like
\begin{equation}
\label{Ein.eqs}
R_{\mu}^{\nu}-\frac12\delta_{\mu}^{\nu}R=
 \kappa\left(T_{\mu(mat)}^{\nu}+T_{\mu(obs)}^{\nu}+T_{\mu(ghost)}^{\nu}\right),
\end{equation}
where $T_{\mu(mat)}^{\nu}$ is the energy-momentum tensor of matter fields, $T_{\mu(obs)}^{\nu}$ and $T_{\mu(ghost)}^{\nu}$ are obtained by varying the gauge-fixing and ghost action, respectively. $T_{\mu(obs)}^{\nu}$ describes the observer (the gauge subsystem) in the extended set of equations.

In particular, the $0\choose 0$-Einstein equation (Hamiltonian constraint) can be presented in the form
\begin{equation}
\label{new_H_constr}
H=E,
\end{equation}
where $H$ is a Hamiltonian in extended phase space and
\begin{equation}
\label{eigenvalue}
E=-\int\sqrt{-g}\,T_{0(obs)}^0\,d^3 x.
\end{equation}
In quantum theory the modified Hamiltonian constraint leads to a stationary Schr\"odinger equation for the physical part of the wave function:
\begin{equation}
\label{stat.states}
H_{(phys)}\Psi_{kn}(Q)=E_n\Psi_{kn}(Q).
\end{equation}
\begin{equation}
\label{stat.WF}
\Psi_k(Q,\,t)=\sum_n c_n\Psi_{kn}(Q)\exp(-iE_n t);
\end{equation}

To give a simple example, in this section we shall bear in mind an
isotropic universe, then the parametrization function $v(\tilde N,Q)$,
as well as the gauge-fixing function $f(Q)$, will depend only on a
scale factor, i. e.
\begin{equation}
\label{isotr.paramet}
\displaystyle\frac{a^3}{N}=v(\tilde N,a),\quad
\tilde N=f(a)+k.
\end{equation}
The quasi-energy-momentum tensor of the gauge subsystem reads:
\begin{equation}
\label{T_obs}
T_{\tilde N(obs)}^{\nu}=
 {\rm diag}\left(\varepsilon_{(obs)},\,
  -p_{(obs)},\,-p_{(obs)},\,-p_{(obs)}\right);
\end{equation}
\begin{equation}
\label{eps_obs}
\varepsilon_{(obs)}=
 -\left.\frac{\dot\pi_0}{2\pi^2}
  \frac{v^2(\tilde N,a)}{a^6 v_{,\tilde N}}\right|_{\tilde N=f(a)+k};
\end{equation}
\begin{equation}
\label{p_obs}
p_{(obs)}=\varepsilon_{(obs)}
 \left.\left[1-\frac{a}{3v(\tilde N,a)}
  \left(v_{,\tilde N}f_{,a}+v_{,a}\right)\right]\right|_{\tilde N=f(a)+k}.
\end{equation}

So, the gauge-fixing term in the action describes a medium with the equation of state depending on the chosen parametrization and gauge. Now, if the gauge variable is the lapse function $N$, and the gauge condition is
\begin{equation}
\label{N-cond}
N=a+\frac1{a^3}.
\end{equation}
the equation of the medium would be
\begin{equation}
\label{state.eq}
p_{(obs)}=\varepsilon_{(obs)}
 \left[1-\frac13\left(\frac4{1+a^4}+2\right)\right].
\end{equation}
In the course of cosmological evolution the equation of state changes from $p_{(obs)}=-\varepsilon_{(obs)}$ in the limit of small $a$ to $p_{(obs)}=\displaystyle\frac13\varepsilon_{(obs)}$ in the limit of large $a$. The former corresponds a medium with negative pressure typical for an exponentially expanded early universe with $\Lambda$-term, the latter is an ultrarelativistic equation of state, and the Einstein equations in the limit of large $a$ have a solution describing a Friedmann universe in the conformal time gauge $N=a$. Therefore, we can see that the gauge subsystem appears to be a factor of cosmological evolution; its state changing over the history of the Universe determining a cosmological scenario.

It is the answer for the third question about the role of gauge degrees of freedom.

\section{Problems}
To summarize, if one rejects the assumption about asymptotic boundary conditions when constructing quantum theory of the whole Universe, one would come to a gauge dependent formulation in which the Schr\"odinger equation for the wave function of the Universe depends on a chosen parametrization and gauge, in other words, on a chosen reference frame, and so do its solutions. The spectrum of the Hamiltonian operator now is not limited by the only line $E=0$, as it takes place in the Wheeler -- DeWitt quantum geometrodynamics, that enables us to solve the problem of time and related problems.

However, as it often happens, every step in research creates new problems. What is the Hilbert space structure for different reference frames? What could be relations between the solutions to the Schr\"odinger equation corresponding to various reference frames? The latter question is of importance, because one can imagine a topologically non-trivial universe, so that one {\it have to} introduce different reference frames in different regions of this universe. Formally, the path integral approach allows us to describe a spacetime various regions of which are considered from the point of view of different reference frames (different observers). Nevertheless, there exist a mathematical problem, how to describe a transition from one reference frame to another.

We can expect that if gauge conditions fixing the reference frames do not differ much from each other, the same is true for corresponding solutions to the Schr\"odinger equation. A possible mathematical task is to find classes of solutions within which the solutions are ``stable'' enough under small variations of gauge conditions. The structure of these classes must be anyhow related with the structure of diffeomorphism group. This task is very laborious since the structure of diffeomorphism group is known to be very complicated. Nevertheless, one can start, as usual, from well-studied subgroups and try to find the way.

An alternative way is to seek for any ``privileged'' reference frame in which the picture of the Universe evolution would better correspond with observational data. However, in my opinion, today we do not have available such significant arguments and do not have any grounds to postulate a privileged reference frame.

In a full description one should also take into account thermodynamical properties of a quantum Universe filled with a medium playing the role of a reference frame. Indeed, one of possible methods to build thermodynamics of the system under consideration is to write a density matrix through a path integral with Euclidean version of an action (in our case it is a gauged gravitational action), so one must expect that thermodynamical properties of the system would depend on a chosen reference frame as well. It must not be surprising for us, since the example of Rindler space teaches us that thermodynamical properties could actually change after going over to another frame. But we yet need a clear interpretation of quantum gravitational phenomena taking place under this transition.

And another important problem is that of the transition to classical limit of General Relativity when all gauge-dependent effects must vanish. In particular, in quantum stage of its evolution the Universe can be found, in general, in any eigenstate of the Hamiltonian operator with a non-zero eigenvalue. We need for some mechanism which would explain how in the result of quantum evolution the Universe appears to be in the state with zero eigenvalue of the Hamiltonian. This mechanism should be general enough not to depend on a chosen model. Only in this case we can reach better understanding of quantum gravitational processes in the Early Universe.

\end{document}